# Self-aligning universal beam coupler

David A. B. Miller

*Ginzton Laboratory, Stanford University, 348 Via Pueblo Mall, Stanford CA 94305-4088, USA*
*dabm@ee.stanford.edu*

**Abstract:** We propose a device that can take an arbitrary monochromatic input beam and couple it into a single-mode guide or beam. Simple feedback loops from detectors to modulator elements allow the device to adapt automatically to any specific input beam form. Potential applications include automatic compensation for misalignment and defocusing of an input beam, coupling of complex modes or multiple beams from fibers or free space to single-mode guides, and retaining coupling to a moving source. Straightforward extensions allow multiple different overlapping orthogonal input beams to be separated simultaneously to different single-mode guides with no splitting loss in principle. The approach is suitable for implementation in integrated optics platforms that offer elements such as phase shifters, Mach-Zehnder interferometers, grating couplers, and integrated monitoring detectors, and the basic approach is applicable in principle to other types of waves, such as microwaves or acoustics.

## 1. Introduction

There has recently been growing interest in exploiting multiple modes, both in optical fibers [1] and free-space [2], for expanding communications bandwidth and capabilities. Selecting and coupling to complicated mode forms such as higher fiber modes [1] or angular momentum beams [2] is challenging, however, especially if splitting losses are to be avoided. Coupling to waveguides generally remains difficult in optics, especially if alignment or precise focusing cannot be guaranteed. Simultaneous coupling of multiple overlapping input modes without splitting loss has had few known solutions [3,4]. Here we propose a novel approach to coupling that both allows complicated modes to be coupled, e.g., to single-mode waveguides, and can accommodate misalignments and movements, all without moving parts. It can also couple multiple different overlapping modes simultaneously to different output waveguides, without fundamental splitting loss. The approach can be implemented using standard integrated optical components, detectors and simple local feedback loops, and is automatic, not requiring advance knowledge of the beam form to be coupled. The approach could be applied in principle to other waves, such as radio waves, microwaves or acoustics [5].

## 2. Device concept

Fig. 1 shows a conceptual schematic of the approach. For simplicity for the moment, we consider a beam varying only in the lateral direction. For illustration we divide the arbitrary input beam into 4 pieces, each incident on a different one of the 4 beam splitter blocks. Each block includes a variable reflector (except number 4, which is 100% reflecting) and a phase shifter. (The phase shifter PS1 is optional, allowing the overall output phase of the beam to be controlled.) We presume loss-less devices whose reflectivity and phase shift can be set independently, for example, by applied voltages for electrooptic or thermal control. For the moment, we neglect diffraction inside the optics and presume that the phase shifters, reflectors, and detectors operate equally on the whole beam going through one beamsplitter.

We shine the input beam onto the beamsplitter blocks as shown. Initially, the phase shifter and reflectivity settings can be arbitrary as long as the reflectivities are non-zero so that we start with non-zero powers on the detectors. First, we adjust the phase shifter P4 to minimize the power on detector D3. Doing so ensures that the wave reflected downwards from beamsplitter 3 is in antiphase with any wave transmitted from the top through beamsplitter 3. Then we adjust the reflectivity R3 to minimize the power in detector D3 again, now completely cancelling the transmitted and reflected beams coming out of the bottom of beamsplitter 3. (If there are small phase changes associated with adjusting reflectivity, then we can iterate this process, adjusting the phase shifter again, then the reflectivity, and so on, to minimize the D3 signal.)

We then repeat this procedure for the next beamsplitter block, adjusting first phase shifter P3 to minimize the power in detector D2, and then reflectivity R2 to minimize the D2 signal again. We repeat this procedure along the line of phase shifters, beamsplitters and detectors. Finally, all the power in the incident beam emerges from the output port on the right. (This approach could also be used to combine multiple beams of unknown relative phases, as in fiber laser systems [7], with each beam incident on a separate beamsplitter block.)

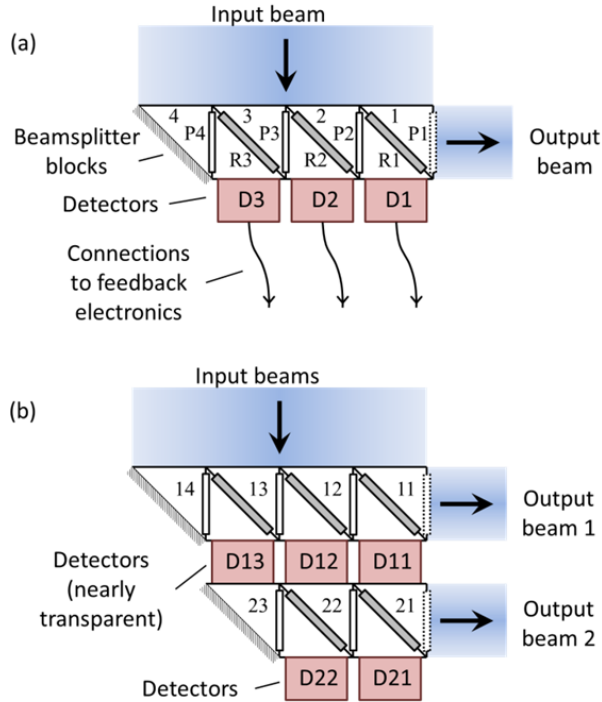

Fig. 1. Schematic illustration of the device structure. Diagonal grey rectangles represent controllable partial reflectors. Vertical clear rectangles represent controllable phase shifters.. (a) Coupler for a single input beam with four beamsplitter blocks (numbered 1 – 4), phase shifters P1 – P4 and reflectors R1 - R3. (b) Coupler for two simultaneous orthogonal input beams (connections from detectors to feedback electronics omitted for clarity).

Unlike typical adaptive optical schemes (see, e.g., Refs.[5,6]), this method is progressive rather than iterative – the process is complete once we have stepped once through setting the elements one by one – and only requires local feedback for minimization on one variable at a time – no global calculation of a merit function or simultaneous multiparameter optimization is required. Simple low-speed electronics could implement the feedback.

To optimize this beam coupling continually, we can leave this feedback system running as we use the device, stepping cyclically through the minimizations as discussed. This would allow real-time tracking and adjustments for misalignments or to retain coupling to moving sources. For static sources, we could use an alternate algorithm based only on maximizing output beam power (see Appendix A).

**3. Waveguide device**

Fig. 2 shows a waveguide version based on Mach-Zehnder interferometers (MZIs) as the adjustable "reflectors" and phase shifters, with Fig. 2 (a) corresponding to Fig. 1 (a). A MZI gives variable overall phase shift of both outputs based on the common mode drive of the controllable phase elements in each arm and variable "reflectivity" (i.e., splitting between the output ports) based on the differential arm drive. Such a waveguide approach avoids diffractions inside the apparatus and allows equal path lengths for all the beam segments. Equal path lengths are important for operation over a broad wavelength range or bandwidth; otherwise the relative propagation phase changes with wavelength in the different waveguide paths.

For further equality of beam paths and losses, we could add dummy MZIs in paths 1, 2, and 3, respectively, as shown in Fig. 2 (b), to give the same number of MZIs in every beam path through the device; the dummy devices would be set so as not to couple between the adjacent waveguides (i.e., the "bar" rather than the "cross" state), and to give a standard phase shift. Note that as long as no power is lost from the system out of the "open" arms – here, the top right ports of the top two dummy devices in Fig. 2 (b) – the settings of these dummy devices are not critical; the subsequent setting of MZIs 1 – 4 can compensate for any such loss-less modification of the input waves. We could add further detectors at those top right ports, adjusting the dummy MZI reflectivities to minimize the signals in such detectors, ensuring loss-less operation. We note that systems with large numbers (e.g., 2048) of MZIs have been demonstrated experimentally, with low overall loss [8], so the relatively large arrays that might be required for complex beams could be feasible practically. An alternative scheme to that of Fig. 2(a) or 2(b) using a binary tree of devices for coupling to a single input beam is presented in Appendix B.

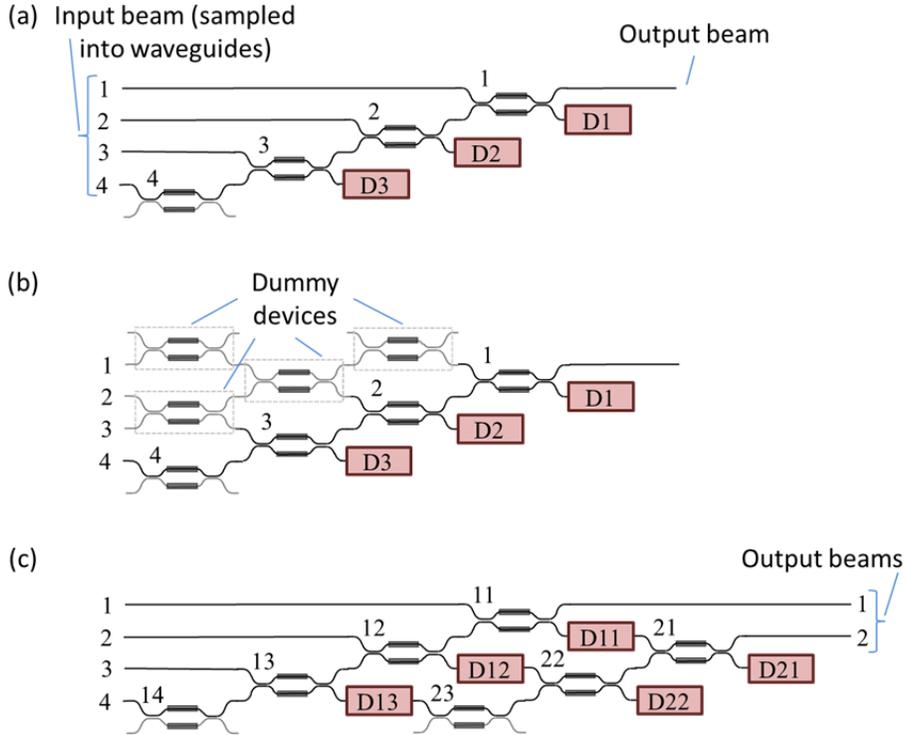

Fig. 2. Mach-Zehnder implementation with detectors. Device numberings correspond to those of Fig. 1. (a) Coupler for a single input beam. (b) Coupler as in (a) with dummy devices added to ensure equal path lengths and background losses. (c) Coupler for two simultaneous modes. The greyed-out lower portions in the bottom row of Mach-Zehnder devices are optional arms for symmetry only; simple controllable phase shifters could be substituted for these Mach-Zehnder devices.

To use the waveguide scheme with a spatially continuous input beam, we need to put the different portions of the beam into the different waveguides. We could use one grating coupler [9] per waveguide as explored for angular momentum beams [3] or phased-array antennas [10]. For full 2D arrays, we need space to pass the waveguides between the grating couplers [9]. We could either allow an imperfect fill factor, shining the whole beam onto the top of the grating coupler array (as in Fig. 3(a)), or we could use an array of lenslets focusing the beam portions onto the grating couplers to improve the fraction of the beam that lands on the grating couplers (Fig. 3 (b)). Grating coupler approaches are also known that can separate polarizations to two separate channels [9,11-13], allowing the input mode of interest to have arbitrary polarization content at the necessary expense of twice as many channels in the device overall.

**4. Separation of multiple orthogonal beam**

We can extend this concept to detecting multiple orthogonal modes simultaneously. In this case, we would use detectors that are mostly transparent, such as silicon defect-enhanced photodetectors in telecommunications wavelength ranges [14-17], sampling only a small amount of the power and transmitting the rest. Now, (Figs. 1(b) and 2(c)), we first set the "top" row of phase shifts and reflections (devices $11 - 14$) as before while shining the first beam on the device, which gives output beam 1. Then, if we shine a second, orthogonal beam on the device, it will transmit completely through the "top" row of beamsplitters and photodetectors, becoming an input beam for the second row.

To understand why this second orthogonal beam passes through the first row, note that, for any loss-less beam coupler that couples all of one input mode (e.g., our first input beam shining into our device) into a single output mode (here output beam 1), it is impossible to combine any power from any other orthogonal beam into the same output mode [18]; since there is nowhere else for the power in some second orthogonal beam to go (nothing in our optics reflects this second beam backwards), it will all pass through to the second row of beamsplitter blocks. The beam form will certainly be changed as it passes through the first, but since our process can adapt to an arbitrary beam, we can simply use the same alignment process as before, now with the second row of phase shifters and reflectors (devices $21 - 23$), and the detectors D21 – D22, to direct all of this power to output beam 2.

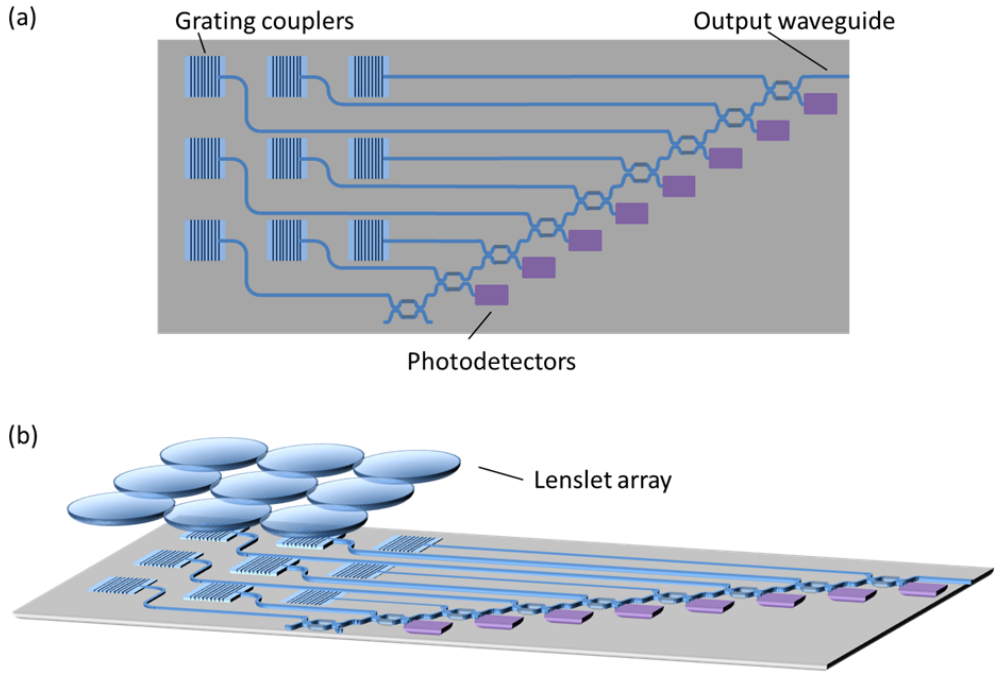

Fig. 3. (a) Figurative top-view schematic of an array of grating couplers with a set of Mach-Zehnder devices to produce one beam at the output waveguide, analogous to Fig. 2(a). For graphic simplicity, we omit here any additional lengths of waveguide and possible dummy Mach-Zehnder devices to equalize path lengths and losses. (b) Illustration of the addition of a lenslet array to improve the fill factor. The input beam is shone onto the grating coupler surface or onto the lenslet array.

We can repeat this process for further orthogonal beams using further rows, up to the point where the number of rows of beamsplitters (and the number of output beams) equals the number of beamsplitter blocks in the first row. (Generally, we can leave all the preceding beams on, if we wish, as we adjust for successive added orthogonal beams.) We could analogously apply the same approach to the structure of Fig. 3, adding further "rows" as in Fig. 2 (c) to allow simultaneous detection of multiple orthogonal 2D beams shone on the grating couplers or lenslets.

We could also add some identifying coding to each orthogonal input beam, such as a small amplitude modulation at a different frequency for each beam; then, we can have all beams on at once, with each detector row looking only for one beam's specific modulation frequency or coding signal. Such an approach, combined with continuous cycling through the different rows as above, allows continuous tracking and alignment adjustment of all the orthogonal beams while leaving all of them on.

## 5. Discussion

The number of portions or subdivisions we need to use for a given beam depends on how complex a mode we want to select or how complicated a correction we want to apply. Generally, this device corresponds to the approach to complexity counting discussed in Ref. [19]. If we want to be able to select one specific input mode form out of $M_I$ orthogonal possibilities, we need at least $M_I$ beamsplitter blocks in the (first) row. Subsequent rows to select other specific modes from this set need, progressively, one fewer beamsplitter block. The number of rows of beamsplitter blocks is the mode coupling number $M_C$ in the notation of Ref. [19]

At radio or microwave frequencies, we could use antennas instead of grating couplers. Various microwave splitters and phase shifters are routinely possible [20]. Use of nanometallic or plasmonic antennas [21], waveguides [22], modulators [23] and detectors [21,22] is also conceivable for subwavelength circuitry in optics, allowing possibly very small and highly functional mode separation and detection schemes.

## 6. Conclusions

In conclusion, we have shown a general method for coupling an arbitrary input beam to one specific output beam, such as a waveguide mode, with an automatic method for setting the necessary coefficients in the array of adjustable reflectors and phase shifters based on signals from photodetectors, and with extensions to allow multiple orthogonal input beams to be separated without fundamental splitting loss. This should open a broad range of flexible and adaptable optical functions and components, with analogous possibilities for other forms of waves such as microwaves and acoustics.

**Appendix A - Alternate alignment algorithm**

As an alternative to the use of multiple detectors when aligning a single beam with the device, we could use only a detector in the output beam, with a different algorithm. We first set all the reflectors in the beam splitter blocks in Fig. 1(a) to be 100% transmitting, except the last one – beamsplitter block 4, which is set permanently to 100% reflection – and the second last one (block 3), which we set to some intermediate value of reflectivity. Then, monitoring a detector in the output beam (port 1 on the right), we adjust the phase shifter P4 on the right of beamsplitter block 4 to maximize the output power. We then adjust the reflectivity R3 in beamsplitter block 3 to maximize the output power again (these two steps in sequence arrange that there is no power emerging from the bottom of block 3). We then proceed along the beamsplitter blocks in a similar fashion, setting the next beamsplitter reflectivity to some initial intermediate value, adjusting the phase shifter just to its left to maximize the output, then adjusting the reflectivity in this block to maximize output again, and so on along the beamsplitter blocks. (In this case, we would not be able to do continuous feedback on the settings while the system was running because we need to set some of the reflectors temporarily to 100% transmission during the optimization steps.)

This approach can be extended to multiple orthogonal beams. Once we have set the first row, we leave those settings fixed and then proceed with aligning the second orthogonal beam in a similar fashion, monitoring the power now in the output beam in the second row.

**Appendix B - Alternate configuration for single beam coupling**

For coupling a single beam, an alternate configuration of phase shifters, MZ interferometers and detectors is shown in Fig. 4. In this approach, phase shifter P1 is adjusted to minimize the signal in detector DA1, and then the split ratio ("reflectivity") of MZI MA1 is adjusted through differential drive of the arms to minimize the DA1 signal again. Similar processes can be used simultaneously with P2, DA2 and MA2, with P3, DA3 and MA3, and with P4, DA4 and MA4. Next, the overall phase is adjusted in MA1 to minimize signal in DB1, and then the split ratio ("reflectivity") of MB1 is adjusted to minimize the DB1 signal again. A similar process can be run simultaneously with MA3, DB2 and MB2. Finally, in this example, the phase in MB1 is adjusted to minimize the DC1 signal, and then the split ratio ("reflectivity") of MC1 is adjusted to minimize the DC1 signal again. Dummy phase shifters can be incorporated in the input paths for beams 2, 4, 6, and 8, as shown to help ensure equality of path lengths in the system overall.

This approach has the advantages of requiring no dummy interferometers and allowing simultaneous feedback loop adjustments, first in the DA column of detectors, then in the DB column, and finally in the DC column. In contrast to the approach of Fig. 2(a) of the main text, the MZI devices are arranged in a binary tree rather than a linear sequence, so the device is shorter and a given beam travels through fewer MZI devices, possibly reducing loss. We could extend this approach also for coupling multiple orthogonal beams (e.g., by using beams transmitted through mostly transparent versions of the detectors into analogous, shorter trees of devices); but, unlike the approaches of Fig. 2, we would require crossing waveguides and/or multiple stacked planar circuits if we used a planar optical approach.

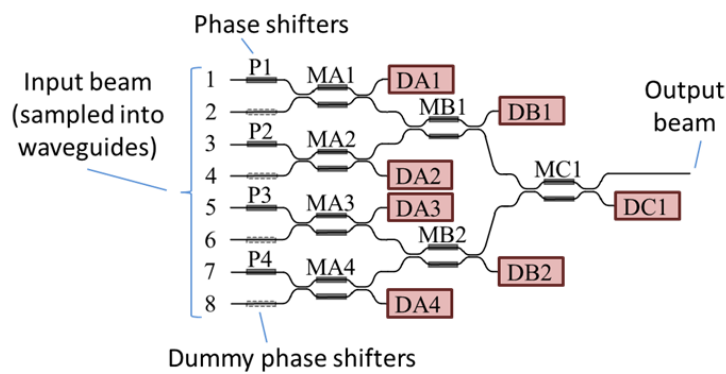

Fig. 4. Alternative binary approach for coupling one arbitrary input beam to a single output beam. P1 – P4 are controllable phase shifters, MA1 – MC1 are controllable MZ interferometers, and DA1 – DC1 are detectors used to give the signals for feedback loops. The dummy phase shifters are optional and could be included for equality of path lengths and/or loss.

**Acknowledgements**

This project was supported by funds from Duke University under an award from DARPA InPho program, and by Multidisciplinary University Research Initiative grants (Air Force Office of Scientific Research, FA9550-10-1-0264 and FA9550-09-0704).

**References**

1. R. Ryf, S. Randel, A. H. Gnauck, C. Bolle, A. Sierra, S. Mumtaz, M. Esmaeelpour, E. C. Burrows, R.-J. Essiambre, P. J. Winzer, D. W. Peckham, A. H. McCurdy, and R. Lingle, Jr., "Mode-Division Multiplexing Over 96 km of Few-Mode Fiber Using Coherent 6x6 MIMO Processing," J. Lightwave Technol. **30**, 521-531 (2012).
2. A. E. Willner, J. Wang, and H. Huang, "A Different Angle on Light Communications," Science **337**, 655-656 (2012).
3. T. Su, R. P. Scott, S. S. Djordjevic, N. K. Fontaine, D. J. Geisler, X. Cai, and S. J. B. Yoo, "Demonstration of free space coherent optical communication using integrated silicon photonic orbital angular momentum devices," Opt. Express **20**, 9396-9402 (2012). http://www.opticsinfobase.org/oe/abstract.cfm?URI=oe-20-9-9396
4. Y. Jiao, S. Fan, D. A. B. Miller, "Demonstration of Systematic Photonic Crystal Device Design and Optimization By Low Rank Adjustments: an Extremely Compact Mode Separator," Optics Letters **30**, 141-143 (2005).
5. A. P. Mosk, A. Lagendijk, G. Lerosey and M. Fink, "Controlling waves in space and time for imaging and focusing in complex media," Nature Photonics 6, 283–292 (2012). doi:10.1038/nphoton.2012.88
6. R. N. Mahalati, D. Askarov, J. P. Wilde, and J. M. Kahn, "Adaptive control of input field to achieve desired output intensity profile in multimode fiber with random mode coupling," Opt. Express **20**, 14321-14337 (2012). http://www.opticsinfobase.org/oe/abstract.cfm?URI=oe-20-13-14321
7. Shay, T. M. Theory of electronically phased coherent beam combination without a reference beam *Opt. Express* **14**, 12188-12195 (2006). http://www.opticsinfobase.org/oe/abstract.cfm?URI=oe-14-25-12188
8. S. Sohma, T. Watanabe, N. Ooba, M. Itoh, T. Shibata, and H. Takahashi, "Silica-based PLC type 32 x 32 optical matrix switch," European Conference on Optical Communications (ECOC 2006), Cannes, France, 24 – 28 Sept. 2006, Paper OThV4. 10.1109/ECOC.2006.4801113
9. G. Roelkens, D. Vermeulen, S. Selvaraja, R. Halir, W. Bogaerts, and D. Van Thourhout, "Grating-Based Optical Fiber Interfaces for Silicon-on-Insulator Photonic Integrated Circuits," IEEE J. Quantum Electron. **17**, 571 – 580 (2011).
10. K. Van Acoleyen, H. Rogier, and R. Baets, "Two-dimensional optical phased array antenna on silicon-on-Insulator," Opt. Express **18**, 13655-13660 (2010). http://www.opticsinfobase.org/oe/abstract.cfm?URI=oe-18-13-13655
11. W. Bogaerts, D. Taillaert, P. Dumon, D. Van Thourhout, R. Baets, and E. Pluk, "A polarization-diversity wavelength duplexer circuit in silicon-on-insulator photonic wires," Opt. Express **15**, 1567-1578 (2007). http://www.opticsinfobase.org/oe/abstract.cfm?URI=oe-15-4-1567
12. G. Roelkens, D. Van Thourhout, and R. Baets, "Silicon-on-insulator ultra-compact duplexer based on a diffractive grating structure," Opt. Express **15**, 10091-10096 (2007). http://www.opticsinfobase.org/oe/abstract.cfm?URI=oe-15-16-10091
13. F. Van Laere, W. Bogaerts, P. Dumon, G. Roelkens, D. Van Thourhout, and R. Baets, "Focusing Polarization Diversity Grating Couplers in Silicon-on-Insulator," J. Lightwave Technol. **27**, 612 – 618 (2009).
14. T. Baehr-Jones, M. Hochberg, and A. Scherer, "Photodetection in silicon beyond the band edge with surface states," Opt. Express 16, 1659-1668 (2008). http://www.opticsinfobase.org/oe/abstract.cfm?URI=oe-16-3-1659
15. J. D. B. Bradley, P. E. Jessop, and A. P. Knights, "Silicon waveguide-integrated optical power monitor with enhanced sensitivity at 1550 nm," Appl. Phys. Lett. **86**, 241103 (2005). http://dx.doi.org/10.1063/1.1947379
16. D. F. Logan, P. E. Jessop, and A. P. Knights, "Modeling defect enhanced detection at 1550 nm in integrated silicon waveguide photodetectors," J. Lightwave Technol. **27**, 930-937 (2009).
17. M. W. Geis, S. J. Spector, M. E. Grein, J. U. Yoon, D. M. Lennon, and T. M. Lyszczarz, "Silicon waveguide infrared photodiodes with >35 GHz bandwidth and phototransistors with 50 AW-1 response," Opt. Express 17, 5193-5204 (2009). http://www.opticsinfobase.org/oe/abstract.cfm?URI=oe-17-7-5193
18. D. A. B. Miller, "All linear optical devices are mode converters," Opt. Express **20**, 23985-23993 (2012). http://www.opticsinfobase.org/oe/abstract.cfm?URI=oe-20-21-23985
19. D. A. B. Miller, "How complicated must an optical component be?" Submitted to J. Opt. Soc. Am. A. arXiv:1209.5499 [physics.optics]
20. K. Chang (ed.), *Handbook of RF/Microwave Components and Engineering* (Wiley, Hoboken, 2003).
21. L. Tang, S. E. Kocabas, S. Latif, A. K. Okyay, D.-S. Ly-Gagnon, K. C. Saraswat and D. A. B. Miller, "Nanometre-Scale Germanium Photodetector Enhanced by a Near-Infrared Dipole Antenna," Nature Photonics **2**, 226 – 229 (2008). doi:10.1038/nphoton.2008.30
22. D.-S. Ly-Gagnon, K. C. Balram, J. S. White, P. Wahl, M. L. Brongersma, and D. A. B. Miller, "Routing and Photodetection in Subwavelength Plasmonic Slot Waveguides," Nanophotonics **1**, 9–16, (2012). DOI: 10.1515/nanoph-2012-0002
23. J. A. Schuller, E. S. Barnard, W. Cai, Y. C. Jun, J. S. White, and M. L. Brongersma, "Plasmonics for extreme light concentration and manipulation," Nature Materials **9**, 193-204 (2010).